\documentclass[letter]{aa}
\usepackage{txfonts}
\usepackage{general}
\usepackage{dcolumn}
\usepackage[version=3]{mhchem}
\graphicspath{{fig/}}

\begin{document}

\newcommand{\rr}{\ensuremath{{\foun/\fifn}}}
\newcommand{\hbmk}{Paper~I}
\newcommand{\corr}[2]{\textbf{\red {#1}}}

\title{Multiple nitrogen reservoirs in a protoplanetary disk at the epoch of comet and giant planet formation}

\author{%
  P.~Hily-Blant$^{1}$
  \and V. Magalhaes de Souza$^2$
  \and J. Kastner$^{3}$
  \and T. Forveille$^{1}$}
\institute{%
        Univ. Grenoble Alpes, CNRS, IPAG, 38000 Grenoble, France
        \email{pierre.hily-blant@univ-grenoble-alpes.fr}
    \and IRAM, 300 rue de la Piscine, Grenoble, France
    \and Chester F. Carlson Center for Imaging Science, School of Physics \& Astronomy, and Laboratory for Multiwavelength Astrophysics,
    Rochester Institute of Technology, 54 Lomb Memorial Drive, Rochester NY 14623 USA
}
\date{}

\abstract{%
        The isotopic ratio of nitrogen measured in primitive Solar System bodies shows a broad range of values, the origin of which remains unknown. One key question is whether these isotopic reservoirs of nitrogen predate the comet formation stage or are posterior to it. Another central question is elucidating the processes that can produce the observed variations in the $^{14}$N/$^{15}$N isotopic ratio. Disks that orbit pre-main-sequence (TTauri) stars provide unique opportunities for observing the chemical content of analogs of the protosolar nebula and therefore for building a comprehensive scenario that can explain the origin of nitrogen in the Solar System and in planet-forming disks. With ALMA, it has become possible to measure isotopic ratios of nitrogen-bearing species in such environments. We present spectrally {and spatially} resolved observations of the hyperfine structure of the 4-3 rotational transition of HCN and its main isotopologs H$^{13}$CN and HC$^{15}$N in the disk orbiting the 8~Myr old TTauri star TW~Hya. The sensitivity allows directly measuring the HCN/H$^{13}$CN and HCN/HC$^{15}$N abundance ratios with minimal assumptions. Averaged spatially over the disks, the ratios are 86$\pm$4 and {223$\pm$21,} respectively. The latter value is significantly lower than the CN/C$^{15}$N ratio of 323$\pm$30 in this disk and thus provides the first evidence that two isotopic reservoirs of nitrogen are present in a disk at the stage of giant planet and comet formation. Furthermore, we find clear evidence for an increase in the ratio of HCN to HC$^{15}$N with radius. The ratio in the outer disk, at 45 au, is 339$\pm$28, in excellent agreement with direct measurements in the local interstellar medium, and with the bulk nitrogen isotopic ratio predicted from galactic evolution calculations. In the comet formation region at $r=20$~au, the ratio is a factor $\approx3$ lower, 121$\pm$11. This radial increase qualitatively agrees with the scenario in which selective photodissociation of N$_2$ is the dominant fractionation process. However, our isotopic ratios and kinetic temperature of the HCN-emitting layers quantitatively disagree with models of nitrogen chemistry in disks.}

\maketitle

\section{Introduction}
\vspace{-.5ex}
The isotopic ratio of nitrogen, \rr, in various bodies of the Solar System (meteorites, comets, planets, etc.) shows the largest variations among the most abundant constituents (carbon and oxygen), with {values} ranging from $\sim$50 in submicron grains that are immersed in a chondrite matrix (so-called hotspots) to 441 in the Sun and Jupiter \citep{bonal2010, marty2011, fouchet2000, hilyblant2013a}. When, where, and how these isotopic reservoirs of nitrogen formed during the evolution from a molecular cloud to planetary systems remains to be determined. Comets provide a special case among primitive cosmomaterials: regardless of the cometary type (short or long period) and of the carrier of nitrogen that is observed in their coma (CN, HCN, \ce{NH2}, \ce{NH3}, \ce{N2}, or NO), the isotopic ratio is consistently found to be $\approx$140 \citep[][hereafter \hbmk, and references therein]{wampfler2018, hilyblant2017}. This is lower by a factor three than the ratio of the bulk proto-Sun, which is 441. This well-established observational fact remains unexplained.
\begin{figure*}[t]
	\sidecaption
	\centering
	\includegraphics[width=0.70\hsize]{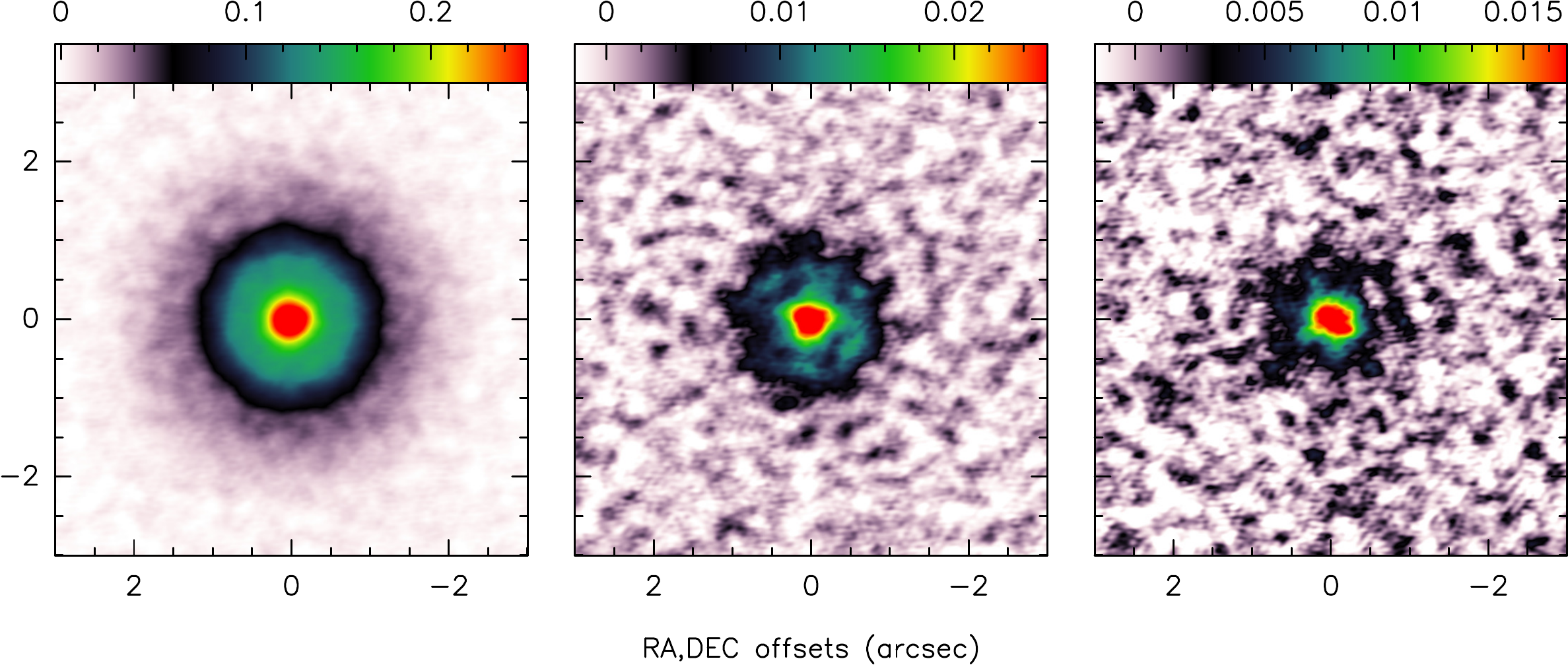}
	\caption{Integrated intensity maps of HCN, H\thcn, and HC\fifn\ (from left to right) in Jy/beam\kms. The maps are centered at $\alpha$=11:01:51.81150, $\delta$=-34:42:17.2636 (J2000.0). The visibilities were projected on the same spatial grid and convolved to a synthesized beam of HPBW $0.26\arcsec\times0.21\arcsec$. The emission is centered on the systemic velocity of 2.83\kms\ and is integrated over 5\kms\ for HCN and 2\kms\ for the less abundant isotopologs.}
	\label{fig:tdv}
\end{figure*}
 
Evolved protoplanetary disks at the stage of giant planet and comet formation provide unique opportunities for investigating {how and when} this \fifn-rich reservoir of nitrogen {was built up}. The nitrogen isotopic ratio was measured via rare isotopologs\footnote{Hereafter, the three isotopologs H$^{12}$C$^{14}$N, $\rm H^{13}C^{14}N$, and $\rm H^{12}C^{15}N$, are denoted HCN, H\thcn, and HC\fifn, respectively.} of HCN, specifically, H\thcn\ and HC\fifn,  for a small sample of disks with the ALMA interferometer \citep{guzman2017}, yielding an HCN/HC\fifn\  between 83$\pm$32 and 156$\pm$71, and a mean value of 111$\pm$19, under the assumption that HCN/H\thcn=70. On the other hand, a direct measurement, one that does not involve assuming a \twc/\thc\ ratio, of CN/C\fifn\ in the TW~Hya system yielded a significantly higher ratio, 323$\pm$30 (\hbmk). Based on this value and on directly measured values in the dense local interstellar medium (ISM), a value of \rr=330$\pm$30 was proposed for the present-day bulk solar neighborhood. This also agrees excellently well with predictions from the galactic chemical evolution model of \cite{romano2017} for the local ISM. A new picture thus emerged in which HCN and CN trace two isotopic reservoirs of nitrogen that are present in analogs of the protosolar nebula (PSN) at the time of comet formation, with HCN representing a \fifn-enriched reservoir of nitrogen similar to that recorded in Solar System comets. However, HCN/HC\fifn\ was only measured indirectly, and moreover, this was not done in TW~Hya, therefore clear-cut evidence for this scenario was lacking.

To elucidate how the various nitrogen reservoirs in the Solar System were formed requires identifying the processes that lead to variations of \rr\  \citep{hilyblant2013a, hilyblant2017, furi2015}. The efficiency of chemical mass fractionation, which is the only process expected in cold and shielded gas \citep{terzieva2000}, has recently been called into question both theoretically \citep{roueff2015, wirstrom2018} and observationally by measuring HCN/HC\fifn\  in the L1498 pre-stellar core \citep{magalhaes2018a}. The obtained value, 338$\pm$28, agrees very well with the present-day elemental ratio of $\sim$330 proposed in \hbmk\ and thus supports the prediction that chemical fractionation is not an efficient process even in cold pre-stellar cores. Further credence was provided by subsequent direct measurements in \ce{HC3N} and \ce{HC5N} toward similar sources \citep{hilyblant2018b, taniguchi2017c}. The concordance between observations and models may be only an apparent one, however, because sharp deviations are recurrently obtained for \ce{N2H+} \citep{bizzocchi2013, redaelli2018}.

The other fractionation process, selective photodissociation of \ce{N2}, which is expected in unshielded gas, has been investigated in the context of protoplanetary disks and molecular clouds \citep{heays2014, visser2018, furuya2018}. The opacity of N\fifn\ is far lower than that of \ce{N2}, therefore more \fifn\ atoms are released upon dissociation than \foun. This leads to an atomic isotopic ratio that is lower than the bulk. Gas-phase and/or surface chemistry then propagates this deviation into other species. A  correlation (positive or negative) between the UV flux and the \rr\ is therefore expected in some species. For HCN in disks, the HCN/HC\fifn\  is predicted to increase with decreasing UV flux.

In this Letter, we aim at providing the requisite evidence for two distinct N reservoirs within the disk that orbits the nearby TTauri star TW~Hya through an analysis of the emission of HCN and isotopologs, based on archival ALMA data (2016.1.00629.S, Cleeves et al). We also explore the possibility of a radial gradient in the HCN/HC\fifn\  to determine whether selective photodissociation is the driving fractionation process in PSN analogs.


\section{Observations and data reduction}
\label{sec:obs}

The (4-3) rotational transitions of HCN, H\thcn, and HC\fifn\ at 354.5, 345.4, and 344.2~GHz, respectively, have been mapped with the ALMA interferometer with a spectral resolution of 61 KHz or 0.053 \kms (see Appendix~\ref{app:obs}). All the lines were observed simultaneously, thus mitigating cross-calibration biases. The standard pipeline of CASA 5.4.0 was used to generate the complex visibilities, which were then self-calibrated in Gildas\footnote{IRAM memo available at \url{http://www.iram.fr/IRAMFR/ARC/documents/filler/casa-gildas.pdf}.}. The final sensitivities are 2.3, 1.9, and 2.0~mJy/beam per 61~kHz channel for HCN, H\thcn, and HC\fifn\ respectively. The resulting integrated intensity maps are shown in \rfig{tdv}.

\begin{figure*}
        \centering
        \sidecaption
        \includegraphics[width=0.69\hsize]{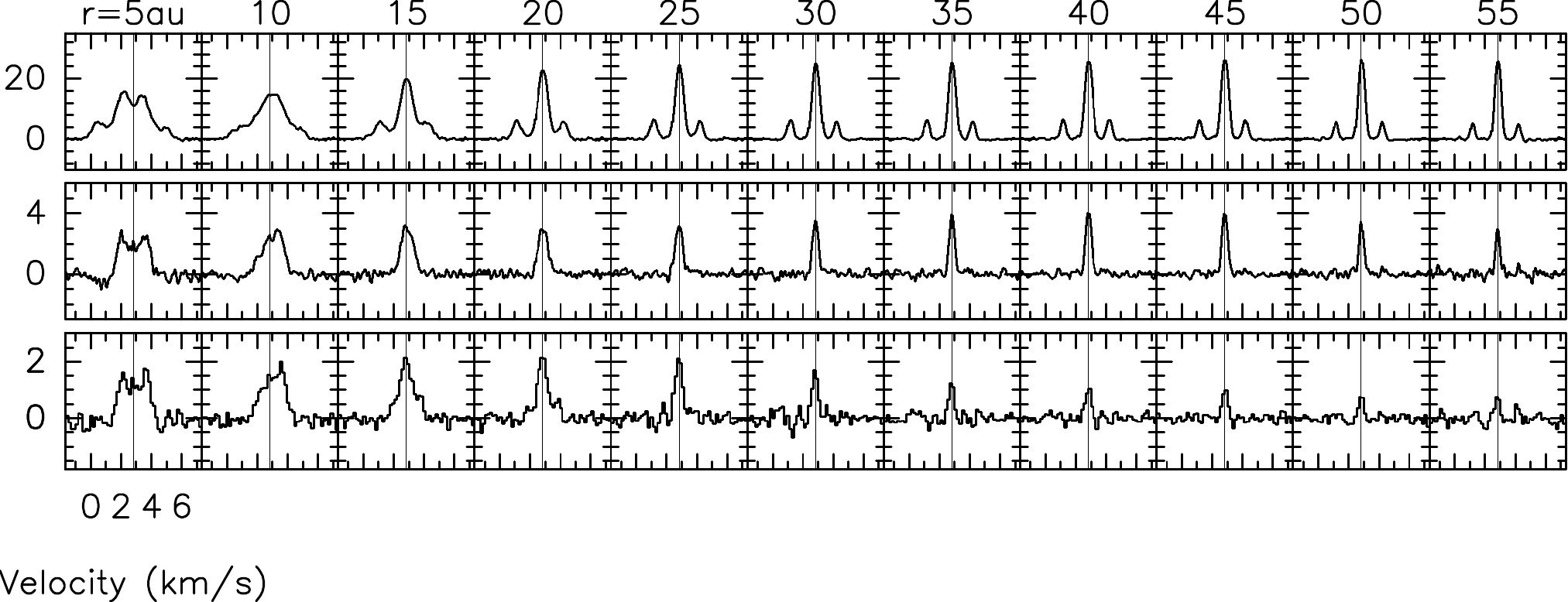}
        \caption{{Spectra of HCN, H\thcn, and HC\fifn\ (from top to bottom) from the Keplerian mask at 2.83 \kms (systemic velocity) and averaged within annuli centered at radii $r=$5 to 55~au. The specific intensity scale is K using K/(Jy/beam) conversion factors of 178.7, 188.1, and 189.0 for HCN, H\thcn, and HC\fifn, respectively.}}
        \label{fig:wspectra}
\end{figure*}

The HCN(4-3) spectra (see \rfig{wspectra}) contain two weak hyperfine (hf) transitions of equal intensity, each carrying 2.1\% of the total flux, and shifted by -1.672 and +1.360\kms\ with respect to our reference frequency 354505.477~MHz. They surround a strong, non-Gaussian, central feature made of three overlapping hf transitions representing 95.8\% of the flux. The intrinsically weaker intensity of H\thcn\ and HC\fifn\ compared to HCN leads to one single feature only, which in the case of H\thcn\  carries 95.8\% of the total flux. For HC\fifn, the splitting is entirely unresolved and the single feature thus carries 100\% of the rotational transition flux (see \rtab{spectro}).

\section{Results}
\label{sec:results}

The conversion of line flux into abundance ratios is a simple procedure provided the lines are optically thin and the excitation temperatures are known. Even for optically thin transitions, however, the overlap in velocity space of lines that are emitted at different locations in the disk, which is a result of the Keplerian rotation of the disk, complicates the conversion from flux into abundance ratios because this mixes different regions of the disk, broadens the lines, and may also weaken the assumption that the emitting species are cospatial. The latter is especially true in the case of species with very different abundances, such as nitrogen isotopologs. Mitigating both the optical depth and Keplerian blurring effects thus offers clear advantages.

The nearly face-on configuration of the TW~Hya disk naturally reduces the Keplerian blurring. In addition, the contribution from different locations along the line of sight to the emerging lines is limited to the vertical direction. Moreover, the spectrally resolved hf multiplet of HCN provides two weak lines that are most likely optically thin. Their flux is thus directly proportional to their column density.

Because the emission of all three isotopologs is well detected and spatially resolved, one goal might be measuring the HCN isotopic ratios as a function of radius, provided that line broadening can be minimized. This implies that the spatial area over which spectra are averaged needs to be limited. Another complementary approach consists of deriving a {spatially averaged isotopic ratio} by measuring line flux ratios over the whole disk. This method provides higher sensitivity, but Keplerian line broadening must still be controlled. We applied both approaches and detail them in the following.

\subsection{Spatially averaged line fluxes}
\label{sec:spatial}

In order to limit the Keplerian line-broadening, the spectra were averaged over several groups of pixels that were selected for their closely similar projected Keplerian velocity. In practice, a 2D mask was thus determined by selecting the disk area where the 354507.455~MHz hyperfine emission of HCN is higher than 40~mJy/beam in each velocity channel. Although the overlap of different Keplerian velocities is well controlled, it is still noticeable close to the disk center, and we further masked out the spectra located at radii smaller than 0.3\arcsec. The resulting mask recovered the usual dipole-like pattern, but with a central hole due to the radial filter (see \rfig{kmask}). As a result, the spectra averaged within each Keplerian velocity mask have a high signal-to-noise ratio (see \rfig{kspectra}), allowing the two weak hf transitions of HCN to be properly fitted. The line fluxes of the two weak hf lines at 354503.869 and 354507.455~MHz were measured through Gaussian fitting and their averaged value, corrected for their summed relative intensity (4.2\%), was then used as an estimator of the HCN flux (see \rtab{kgauss}).

\subsection{Radially dependent fluxes}
\label{sec:radial}
In order to measure the flux ratios as a function of distance from the central TTauri star and still avoid Keplerian line-broadening, a radial mask was superimposed on the previously described Keplerian masks. {To calculate disk radial offsets in au, we adopted a distance of 59.5$\pm$0.9 pc to TW~Hya \citep{gaia2018}}. Spectra contained within concentric annuli, {centered at radii $r=5$ to 55~au in steps of 5~au (one-third of the half-power beam width, HPBW), and of constant thickness $\delta r=$5~au,}  were averaged. The outer radius was imposed by the HC\fifn\ detection limit (see \rfig{tdv}). The number of spectra in each annulus increased with $r$, which compensated for the natural decrease of the signal-to-noise ratio with radius. In the process, spectra were weighted according to their flux rms. As in the spatially averaged approach, the line fluxes of HCN, H\thcn, and HC\fifn\  were measured through Gaussian fitting.

{Radially averaged spectra corresponding to the Keplerian mask at 2.83\kms\ are shown in \rfig{wspectra}. The velocity confusion at small radii is evident, while at radii above 20~au, it is efficiently removed by our masking procedure. Average spectra in all Keplerian channels are displayed in \rfig{spectra_r_vk}. While the peak intensity of HCN hardly changes beyond 20~au, the peak intensities of H\thcn\ and HC\fifn\ decrease when moving outward. Also, all lines become narrower as the radius increases (see \rtab{fwhm}). For HCN, the FWHM is 1.5$\pm0.7$~\kms\ at $r\le0.2\arcsec$, decreasing to 0.34$\pm$0.02\kms\ at 0.9\arcsec.}

\section{Isotopic ratios}
\label{sec:ratio}
\begin{figure*}
        \sidecaption
        \centering
        \includegraphics[height=.34\hsize]{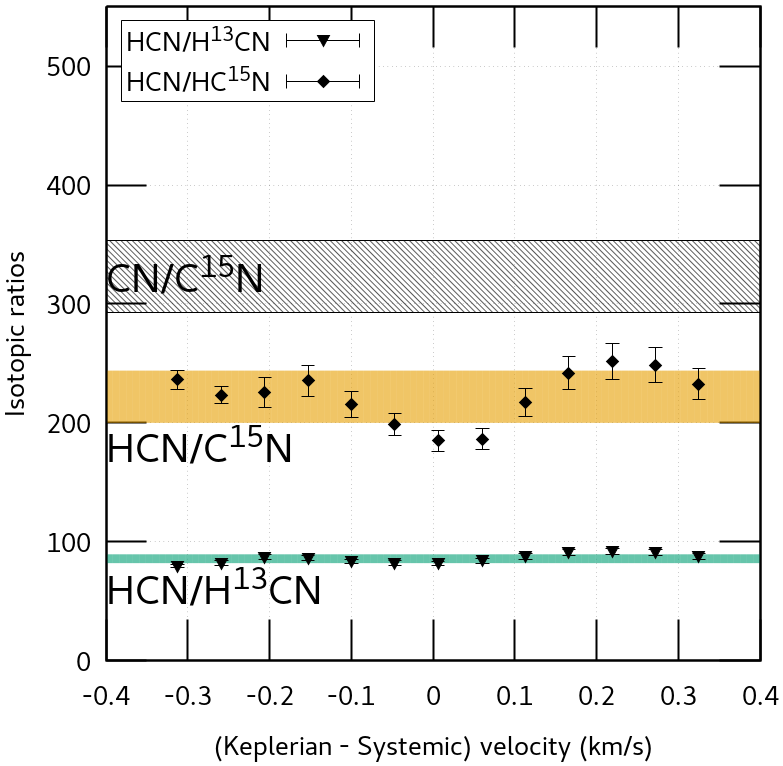}
        \includegraphics[height=.34\hsize]{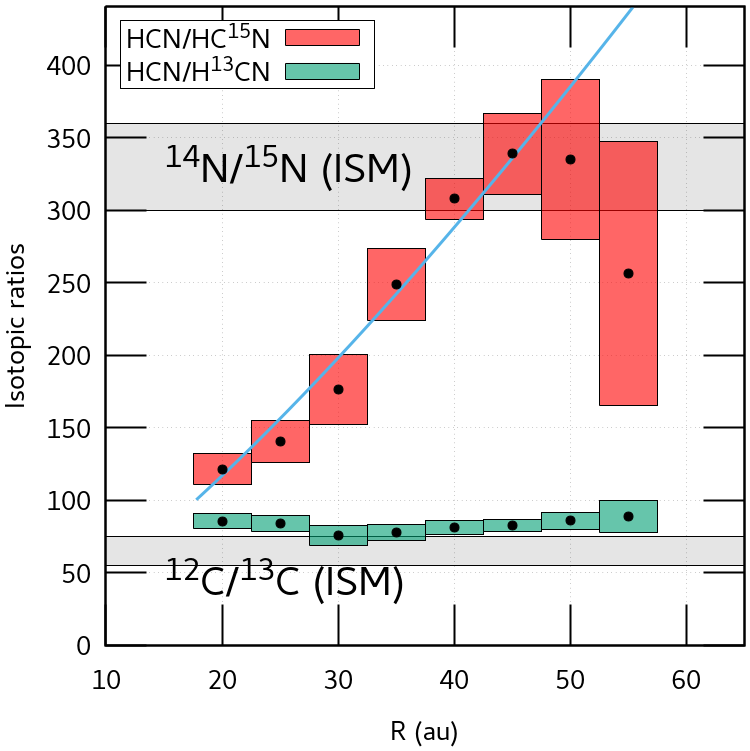}
        \caption{{\bf Left:} HCN/HC\fifn\ and HCN/H\thcn\ isotopic ratios {in each Keplerian channel} derived from the spatially averaged procedure (see \rsec{spatial}). The colored bands indicate the average value $\pm1\sigma$. The spatially averaged CN/C\fifn\ ratio from \hbmk\ is shown as a hatched rectangle. {\bf Right:} {Radial dependence (see \rsec{radial}) of the HCN/H\thcn\ and HCN/HC\fifn\ ratios. Each box is 5~au wide (one-third of the HPBW) and $\pm1\sigma$ in height. The full line shows the best fit of a power law to the \fifn\ ratio, with a slope of 1.3. The gray shaded areas are the present-day \rr\ and \twc/\thc\ elemental ratios in the local ISM (330$\pm$30 (\hbmk) and 65$\pm$10 \citep{halfen2017}, respectively).}}
        \label{fig:wratios}
\end{figure*}

\begin{table}
        \caption{Isotopic ratios derived from our two approaches, the spatially averaged and the radially dependent methods (see also \rfig{wratios}).}
        \label{tab:ratios}
        \centering
        \begin{tabular}{r ccc}
                \toprule
                &HCN/HC\fifn & HCN/H\thcn & H\thcn/HC\fifn \\
                \midrule
                Average                 &223$\pm$21     & 86$\pm$4 & 2.6$\pm$0.2 \\
                \midrule
$r$ (au)=20 & 121.4$\pm$10.6 & 85.6$\pm$ 5.2 & 1.4$\pm$0.1\\
25 & 140.4$\pm$14.6 & 84.2$\pm$ 5.7 & 1.7$\pm$0.1\\
30 & 176.4$\pm$24.1 & 75.9$\pm$ 7.0 & 2.4$\pm$0.3\\
35 & 248.7$\pm$24.6 & 77.9$\pm$ 5.5 & 3.2$\pm$0.4\\
40 & 307.8$\pm$14.2 & 81.2$\pm$ 4.8 & 3.7$\pm$0.3\\
45 & 338.7$\pm$28.2 & 82.5$\pm$ 4.2 & 4.2$\pm$0.4\\
50 & 334.9$\pm$55.0 & 85.9$\pm$ 5.9 & 3.9$\pm$0.7\\
55 & 256.4$\pm$91.2 & 88.7$\pm$11.1 & 2.9$\pm$1.1\\
                \bottomrule
        \end{tabular}
\end{table}

Because the three molecules have very similar spectroscopic constants, their rotational flux ratios are equal to their column density ratios to within 1\%, provided the lines are optically thin and have the same excitation temperature. Under these assumptions, which we verify below, we obtain the isotopic ratios directly from the fluxes. The results are shown in \rfig{wratios} and are listed in Tables~\ref{tab:ratios} and \ref{tab:kratios}.

For HCN/HC\fifn, we find ratios between 185 to 252 across the Keplerian channels (typical uncertainty of 5-10\%) with an average isotopic ratio $223\pm21$. {Here and elsewhere, uncertainties on the absolute flux include a conservative 10\% error (statistical) from amplitude calibration. The quoted uncertainty on average values is the largest of the weighted uncertainties and the dispersion of the averaged values.} The lowest values (185, 186, and 198) are obtained in the three central Keplerian channels, and the average ratio in the eight remaining channels is {233$\pm$12}. HCN/H\thcn\ is more uniform, with values ranging from 82 to 92, which averages to 86$\pm$4. The corresponding average of H\thcn/HC\fifn\  is 2.6$\pm$0.2, with values ranging from 2.22 to 2.75. The low values of HCN/HC\fifn\  might arise because the opacity of HCN is substantial even in its weak hf lines. However,  the almost constant HCN/H\thcn\ suggests that these lines are instead optically thin. Still, assuming a constant isotopic ratio of {233} and adopting the usual expression $\tau/(1-\exp(-\tau))$ for the opacity correction for the integrated intensity, with $\tau$ the opacity \citep{goldsmith1999}, {an opacity of 0.4 for the weakest hf lines, corresponding to a rotational opacity $\tau_{12}\approx 20$, would be needed to explain the flux ratios in these three channels}. Alternatively, assuming a constant H\thcn/HC\fifn\ ratio of 2.7 based on the outer channels (see \rtab{kratios}), the total H\thcn\ opacity would be $\tau_{13}=0.3$ in the three middle channels. It is thus likely that the opacity of the weakest hf line of HCN(4-3) and of H\thcn(4-3) is $\approx 0.3$, which is not entirely negligible, but still fulfills our optically thin assumption. We note that in these three channels, the HCN/H\thcn\ would change by less than 4\% if these corrections were applied.

The radially dependent approach shows that the HCN-to-HC\fifn\ isotopic ratio varies with radius while HCN/H\thcn\ remains constant within 1$\sigma$. More specifically, the former ratio increases from 121 at 20~au to 339 at 45~au, and may decrease at larger distances. The nitrogen isotopic ratio in HCN close to the star is thus a factor {2.7} lower than the spatially averaged CN/C\fifn, 323$\pm$30, whereas {between 45 to 50}~au, the two ratios become consistent within 1$\sigma$. The ratios change by less than 4\% when the three central Keplerian channels are excluded. The best fit of a power law to the HCN/HC\fifn\ gives 117$\pm$10$\times(r/20\,\au)^{1.3\pm0.13}$, indicating a shallower dependence with radius than a pure geometrical dilution of the UV flux from the central star ($\propto r^{-2}$).

A key assumption in our analysis is that a single excitation temperature characterizes the weak hf lines of HCN and the overlapping hf lines of H\thcn\ and of HC\fifn. However, line overlap may lead to unequal excitation temperatures among the hf manifold of HCN \citep{magalhaes2018a}. We have tested our assumption (see Appendix~\ref{app:tex}) by computing the excitation temperatures of HCN, H\thcn, and HC\fifn\ lines while taking line overlap in the excitation into account. Our calculations assume uniform physical conditions and cover the density and gas temperature regimes from \dix{6} to \dix{11}\ccc\ and 5 to 30~K, respectively. We found that deviations from a single excitation temperature are only obtained at densities below 3\tdix{7}\ccc\ (see \rfig{tex}). On the other hand, the full width at half-maximum (FWHM) of the HCN, H\thcn, and HC\fifn\ lines at a radius of 0.6\arcsec\ ($\approx$35~au) and beyond is $\approx$0.35\kms, which places an upper limit on the kinetic temperature of $(FWHM/2.35)^2\times\mu \mh/(2k_B)\approx$35~K \citep{teague2016}. Referring to the TW~Hya model of \cite{vanthoff2017}, this upper limit indicates that the density is at these radii higher than \dix{9}\ccc, which means that it is in a regime where lines are expected to be thermalized. This validates our basic assumption. Our upper limit on the kinetic temperature is also well below the gas temperature of $\sim200$~K where the HCN abundance is the highest in the generic models of \cite{visser2018}.

\section{Discussion}

The spatially averaged directly measured CN/C\fifn\ and HCN/HC\fifn\  in the TW~Hya circumstellar disk provide clear-cut confirmation of at least two isotopic reservoirs of nitrogen, as anticipated by \hbmk. In CN, the ratio is 1.4$\pm$0.2 times higher than in HCN, and their difference is 90$\pm$36. Based on the $\sim$8~Myr age of TW~Hya, these results indicate that multiple isotopic reservoirs are present at the key protoplanetary disk evolutionary stages when comets and giant planets are likely forming.

{In the comet formation zone between 20 and 30~au, the HCN/HC\fifn\  is 121$\pm$11, which is a factor of 2.8 compared to the outer disk at 55~au, where the isotopic ratio of HCN matches the revised elemental ratio of 330 well. The \fifn-enrichment in the region inside of 30~au is reminiscent of the threefold \fifn-enrichment of comets compared to the bulk proto-Sun, providing a strong indication that HCN traces the fractionated reservoir that is recorded by comets. \cite{guzman2017} also found marginal evidence for an increase of the HCN-based \rr\ ratio with radius in the inner $\sim$50~au of the V4046~Sgr disk. We also note that the inner disk ratio confirms the mean HCN/HC\fifn\ ratio of 111$\pm$19 (\hbmk) that was obtained indirectly by \cite{guzman2017} assuming HCN/H\thcn=70$\pm20$. Adopting our measured averaged of 86 for the latter would bring the former ratio to 136$\pm$23. We also very tentatively observe a decrease of the ratio beyond 55~au, which could indicate chemical processes in a gas not exposed to strong UV radiation and where depletion is limited. Measurements with a high signal-to-noise ratio beyond 55~au are needed to confirm this behavior, however.}

The average HCN/H\thcn, $86\pm4$, is significantly higher than the \twc/\thc\ ratio of $65\pm10$ in the local ISM \citep{halfen2017}, but agrees surprisingly well with the value of 89$\pm1$ in the PSN \citep{clayton2004}. This {indicates that HCN is slightly depleted in \thc\ in TW~Hya, opposite to what} was found in the L1498 pre-stellar core \citep{magalhaes2018a}, suggesting that the underlying process may not be the same in these two environments. The depletion of HCN in \thc\ is at variance with the models of \cite{visser2018}, however, which instead predict that HCN is enriched in \thc\ at radii below $\sim$100~au. Optically thick H\thcn\ emission could artificially increase the HCN/H\thcn\ provided that the weak HCN hf transition remains thin. However, the H\thcn\ opacity required to decrease the ratio to the elemental value is 0.55, which is higher than the value of 0.3 derived previously for the central Keplerian channels. This also implies that the total opacity of HCN would be 35, or 0.75 for the weakest HCN hf, which is twice higher than the value of 0.3 that is required to bring the spatially averaged HCN/HC\fifn\  to a constant value (see Sect.~\ref{sec:ratio}). We note that high \twc/\thc\ ratios have already been measured in carbon monoxide from ices toward young stellar objects \citep{smith2015}.

The increasing HCN/HC\fifn\ with radius may indicate that selective photodissociation represents the dominant fractionation process in disks \citep{visser2018}. In these models, the HCN-to-HC\fifn\ column density ratio first decreases with radius before it increases back to the elemental ratio as a consequence of UV flux attenuation that is due to a combination of geometrical dilution and dust extinction. We note that the minimum ratio in the grid of models of \cite{visser2018} is $\approx$240--280 for disks around TTauri stars, which would translate into 180--210 when 330 (rather than 441) is assumed for the elemental ratio. This is still higher than the lowest ratio reported here and might indicate that the UV flux is underestimated in these models. However, dedicated models of TW~Hya are required because the precise radial dependence of the isotopic ratio strongly depends on the physical parameters of the model (disk mass, grain size, and dust-to-gas mass ratio).

Overall, the increasing HCN/HC\fifn\  from 121 up to the bulk value in the outer disk supports a scenario in which a single reservoir of interstellar, not fractionated, nitrogen dominates at large radii in PSN analogs. This nitrogen reservoir is enriched in \fifn\ by the selective photodissociation of \ce{N2} at smaller distance from the central star.

\begin{acknowledgements}
        PHB thanks D. Petry for his help in processing the ALMA data, and A. Faure for useful comments and discussions. We also thank the referee and the editor for valuable comments that greatly clarified the paper. This study makes use of the ADS/JAO.ALMA\#2016.1.00629.S data. ALMA is a partnership of ESO (representing its member states), NSF (USA) and NINS (Japan), together with NRC (Canada), MOST and ASIAA (Taiwan), and KASI (Republic of Korea), in cooperation with the Republic of Chile. The Joint ALMA Observatory is operated by ESO, AUI/NRAO and NAOJ. This work has received financial support from the Institut Universitaire de France and from Labex OSUG@2020. JHK's research on protoplanetary disks orbiting young stars near Earth is supported by NASA Exoplanets      Research Program grants NNX16AB43G and 80NSSC19K0292 to RIT. This study made use of the CDMS database \citep{muller2005}.
\end{acknowledgements}

\bibliographystyle{aa}
\bibliography{general,chemistry}

\begin{thebibliography}{30}
\expandafter\ifx\csname natexlab\endcsname\relax\def\natexlab#1{#1}\fi

\bibitem[{{Beckwith} \& {Sargent}(1993)}]{beckwith1993}
{Beckwith}, S. V.~W. \& {Sargent}, A.~I. 1993, \apj, 402, 280

\bibitem[{{Bizzocchi} {et~al.}(2013){Bizzocchi}, {Caselli}, {Leonardo}, \&
  {Dore}}]{bizzocchi2013}
{Bizzocchi}, L., {Caselli}, P., {Leonardo}, E., \& {Dore}, L. 2013, \aap, 555,
  A109

\bibitem[{{Bonal} {et~al.}(2010){Bonal}, {Huss}, {Krot}, {Nagashima}, {Ishii},
  \& {Bradley}}]{bonal2010}
{Bonal}, L., {Huss}, G.~R., {Krot}, A.~N., {et~al.} 2010, \gca, 74, 6590

\bibitem[{{Clayton} \& {Nittler}(2004)}]{clayton2004}
{Clayton}, D.~D. \& {Nittler}, L.~R. 2004, \araa, 42, 39

\bibitem[{{Daniel} \& {Cernicharo}(2008)}]{daniel2008}
{Daniel}, F. \& {Cernicharo}, J. 2008, \aap, 488, 1237

\bibitem[{{Fouchet} {et~al.}(2000){Fouchet}, {Lellouch}, {B{\'e}zard},
  {Encrenaz}, {Drossart}, {Feuchtgruber}, \& {de Graauw}}]{fouchet2000}
{Fouchet}, T., {Lellouch}, E., {B{\'e}zard}, B., {et~al.} 2000, \icarus, 143,
  223

\bibitem[{{F{\"u}ri} \& {Marty}(2015)}]{furi2015}
{F{\"u}ri}, E. \& {Marty}, B. 2015, Nature Geosc., 8, 515

\bibitem[{{Furuya} \& {Aikawa}(2018)}]{furuya2018}
{Furuya}, K. \& {Aikawa}, Y. 2018, \apj, 857, 105

\bibitem[{{Gaia Collaboration}(2018)}]{gaia2018}
{Gaia Collaboration}. 2018, \aap, 616, A1

\bibitem[{{Goldsmith} \& {Langer}(1999)}]{goldsmith1999}
{Goldsmith}, P.~F. \& {Langer}, W.~D. 1999, \apj, 517, 209

\bibitem[{{Guzm{\'a}n} {et~al.}(2017){Guzm{\'a}n}, {{\"O}berg}, {Huang},
  {Loomis}, \& {Qi}}]{guzman2017}
{Guzm{\'a}n}, V.~V., {{\"O}berg}, K.~I., {Huang}, J., {Loomis}, R., \& {Qi}, C.
  2017, \apj, 836, 30

\bibitem[{{Halfen} {et~al.}(2017){Halfen}, {Woolf}, \& {Ziurys}}]{halfen2017}
{Halfen}, D.~T., {Woolf}, N.~J., \& {Ziurys}, L.~M. 2017, \apj, 845, 158

\bibitem[{{Heays} {et~al.}(2014){Heays}, {Visser}, {Gredel}, {Ubachs}, {Lewis},
  {Gibson}, \& {van Dishoeck}}]{heays2014}
{Heays}, A.~N., {Visser}, R., {Gredel}, R., {et~al.} 2014, \aap, 562, A61

\bibitem[{{Hily-Blant} {et~al.}(2013){Hily-Blant}, {Bonal}, {Faure}, \&
  {Quirico}}]{hilyblant2013a}
{Hily-Blant}, P., {Bonal}, L., {Faure}, A., \& {Quirico}, E. 2013, \icarus,
  223, 582

\bibitem[{{Hily-Blant} {et~al.}(2018){Hily-Blant}, {Faure}, {Vastel},
  {Magalhaes}, {Lefloch}, \& {Bachiller}}]{hilyblant2018b}
{Hily-Blant}, P., {Faure}, A., {Vastel}, C., {et~al.} 2018, \mnras, 480, 1174

\bibitem[{{Hily-Blant} {et~al.}(2017){Hily-Blant}, {Magalhaes}, {Kastner},
  {Faure}, {Forveille}, \& {Qi}}]{hilyblant2017}
{Hily-Blant}, P., {Magalhaes}, V., {Kastner}, J., {et~al.} 2017, \aap, 603, L6

\bibitem[{{Magalh{\~a}es} {et~al.}(2018){Magalh{\~a}es}, {Hily-Blant}, {Faure},
  {Hernandez-Vera}, \& {Lique}}]{magalhaes2018a}
{Magalh{\~a}es}, V.~S., {Hily-Blant}, P., {Faure}, A., {Hernandez-Vera}, M., \&
  {Lique}, F. 2018, \aap, 615, A52

\bibitem[{{Marty} {et~al.}(2011){Marty}, {Chaussidon}, {Wiens}, {Jurewicz}, \&
  {Burnett}}]{marty2011}
{Marty}, B., {Chaussidon}, M., {Wiens}, R.~C., {Jurewicz}, A.~J.~G., \&
  {Burnett}, D.~S. 2011, Science, 332, 1533

\bibitem[{{M{\"u}ller} {et~al.}(2005){M{\"u}ller}, {Schl{\"o}der}, {Stutzki},
  \& {Winnewisser}}]{muller2005}
{M{\"u}ller}, H.~S.~P., {Schl{\"o}der}, F., {Stutzki}, J., \& {Winnewisser}, G.
  2005, Journal of Molecular Structure, 742, 215

\bibitem[{{Redaelli} {et~al.}(2018){Redaelli}, {Bizzocchi}, {Caselli}, {Harju},
  {Chac{\'o}n-Tanarro}, {Dore}, \& {Leonardo}}]{redaelli2018}
{Redaelli}, E., {Bizzocchi}, L., {Caselli}, P., {et~al.} 2018, ArXiv e-prints
  [\eprint[arXiv]{1806.01088}]

\bibitem[{{Romano} {et~al.}(2017){Romano}, {Matteucci}, {Zhang},
  {Papadopoulos}, \& {Ivison}}]{romano2017}
{Romano}, D., {Matteucci}, F., {Zhang}, Z.-Y., {Papadopoulos}, P.~P., \&
  {Ivison}, R.~J. 2017, \mnras, 470, 401

\bibitem[{{Roueff} {et~al.}(2015){Roueff}, {Loison}, \& {Hickson}}]{roueff2015}
{Roueff}, E., {Loison}, J.~C., \& {Hickson}, K.~M. 2015, \aap, 576, A99

\bibitem[{Smith {et~al.}(2015)Smith, Pontoppidan, Young, \& Morris}]{smith2015}
Smith, R.~L., Pontoppidan, K.~M., Young, E.~D., \& Morris, M.~R. 2015, The
  Astrophysical Journal, 813, 120

\bibitem[{{Taniguchi} \& {Saito}(2017)}]{taniguchi2017c}
{Taniguchi}, K. \& {Saito}, M. 2017, \pasj, 69, L7

\bibitem[{{Teague} {et~al.}(2016){Teague}, {Guilloteau}, {Semenov}, {Henning},
  {Dutrey}, {Pi{\'e}tu}, {Birnstiel}, {Chapillon}, {Hollenbach}, \&
  {Gorti}}]{teague2016}
{Teague}, R., {Guilloteau}, S., {Semenov}, D., {et~al.} 2016, \aap, 592, A49

\bibitem[{{Terzieva} \& {Herbst}(2000)}]{terzieva2000}
{Terzieva}, R. \& {Herbst}, E. 2000, \mnras, 317, 563

\bibitem[{{van't Hoff} {et~al.}(2017){van't Hoff}, {Walsh}, {Kama}, {Facchini},
  \& {van Dishoeck}}]{vanthoff2017}
{van't Hoff}, M.~L.~R., {Walsh}, C., {Kama}, M., {Facchini}, S., \& {van
  Dishoeck}, E.~F. 2017, \aap, 599, A101

\bibitem[{{Visser} {et~al.}(2018){Visser}, {Bruderer}, {Cazzoletti},
  {Facchini}, {Heays}, \& {van Dishoeck}}]{visser2018}
{Visser}, R., {Bruderer}, S., {Cazzoletti}, P., {et~al.} 2018, \aap, 615, A75

\bibitem[{{Wampfler} {et~al.}(2018){Wampfler}, {Rubin}, {Altwegg},
  {J{\o}rgensen}, {Calcutt}, \& {Coutens}}]{wampfler2018}
{Wampfler}, S., {Rubin}, M., {Altwegg}, K., {et~al.} 2018, in COSPAR Meeting,
  Vol.~42, 42nd COSPAR Scientific Assembly, B1.3--21--18

\bibitem[{{Wirstr{\"o}m} \& {Charnley}(2018)}]{wirstrom2018}
{Wirstr{\"o}m}, E.~S. \& {Charnley}, S.~B. 2018, \mnras, 474, 3720

\end{thebibliography}

\clearpage
\newpage
\appendix
\section{Observations}
\label{app:obs}

The observations were performed with the ALMA interferometer (2016.1.00629.S, PI:I. Cleeves) and were spread over six epochs (2016 December 30, 2017 July 5, 2017 July 10, 2017 July 15, 2017 July 21, and 2017 July 22). J1037-2934 was used as a bandpass, flux, and phase calibrator. The phase was also calibrated with J1103-3251, and J1058+0133 was also used for bandpass calibration.

H\thcn\ and HC\fifn\ were tuned in the lower side-band and HCN in the upper side-band, allowing simultaneous observations for the three species. We used the standard pipeline of CASA 5.4.0 to generate the first sets of complex visibilities, which were then exported into the Gildas format using the Gildas-Casa filer of Guilloteau, Chapillon et al. (see the IRAM memo available at \url{http://www.iram.fr/IRAMFR/ARC/documents/filler/casa-gildas.pdf}). Self-calibration was then applied to generate the final set of visibilities.

Spectroscopy shows that the three hyperfine (hf) transitions of HCN(4-3) are separated by less than 0.13\kms\ and are not resolved. Their total relative intensity is 95.8\%. The hf transition at 354505.846~MHz is not detected given its 0.03\% relative strength.  The properties of the hf lines of HCN and H\thcn\ are summarized in \rtab{spectro}, while for HC\fifn, the unresolved hyperfine splitting is not given. We also show in \rfig{kmask} channel maps of HCN strong central feature, H\thcn, and HC\fifn. The velocity range covers the main and overlapping hf lines of HCN.

\begin{table}[h]
        \caption{Spectroscopic information on the observed transitions.}
        \label{tab:spectro}
        \centering
        \begin{tabular}{lcrccc}
                \toprule
                Rest Freq.$^a$ & \(\delta v\)$^b$ & \aul$^c$ & $g_u$$^d$ & R. I.$^e$\\
                MHz & \kms & \pers \\
                \midrule
                \mc{5}{HCN}\\
                354503.869 &  1.360 & 1.284\,(-4) & 9 & 2.083\,(-2)\\
                354505.367 &  0.094 & 1.886\,(-3) & 7 & 2.381\,(-1)\\
                354505.477 &  0.000 & 1.925\,(-3) & 9 & 3.125\,(-1)\\
                354505.523 & -0.039 & 2.054\,(-3) &11 & 4.075\,(-1)\\
                354505.846 & -0.312 & 2.620\,(-6) & 7 & 3.307\,(-4)\\
                354507.455 & -1.672 & 1.650\,(-4) & 7 & 2.083\,(-2)\bigskip\\
                \mc{5}{H\thcn}\\
                345338.160&  -1.398&   1.190(-4)&   9&   2.089(-2)\\
                345339.660&  -0.095&   1.740(-3)&   7&   2.376(-1)\\
                345339.770&   0.000&   1.780(-3)&   9&   3.125(-1)\\
                345339.810&   0.035&   1.900(-3)&  11&   4.077(-1)\\
                345340.140&   0.321&   2.420(-6)&   7&   3.305(-4)\\
                345341.750&   1.719&   1.530(-4)&   7&   2.089(-2)\bigskip\\
                \mc{5}{HC\fifn}\\
                344200.109&   0.000&    1.8793(-3) & 9 & 1 \\
                \bottomrule
        \end{tabular}
        \tabnote Numbers are written in the form $a(b) = a\tdix{b}$.\\
        $^a$ Rest frequency.
        $^b$ $\delta v$ is the velocity shift of each hf transition within a given multiplet.\\
        $^c$ Coefficient for spontaneous decay.\\
        $^d$ Upper level total degeneracy.\\
        $^e$ The relative intensity of each hf line, normalized to a sum of 1.
\end{table}

\section{Keplerian masks}
\label{app:kmask}

\subsection{Results from the spatially averaged procedure}
\begin{figure*}
        \centering
        \includegraphics[width=\hsize]{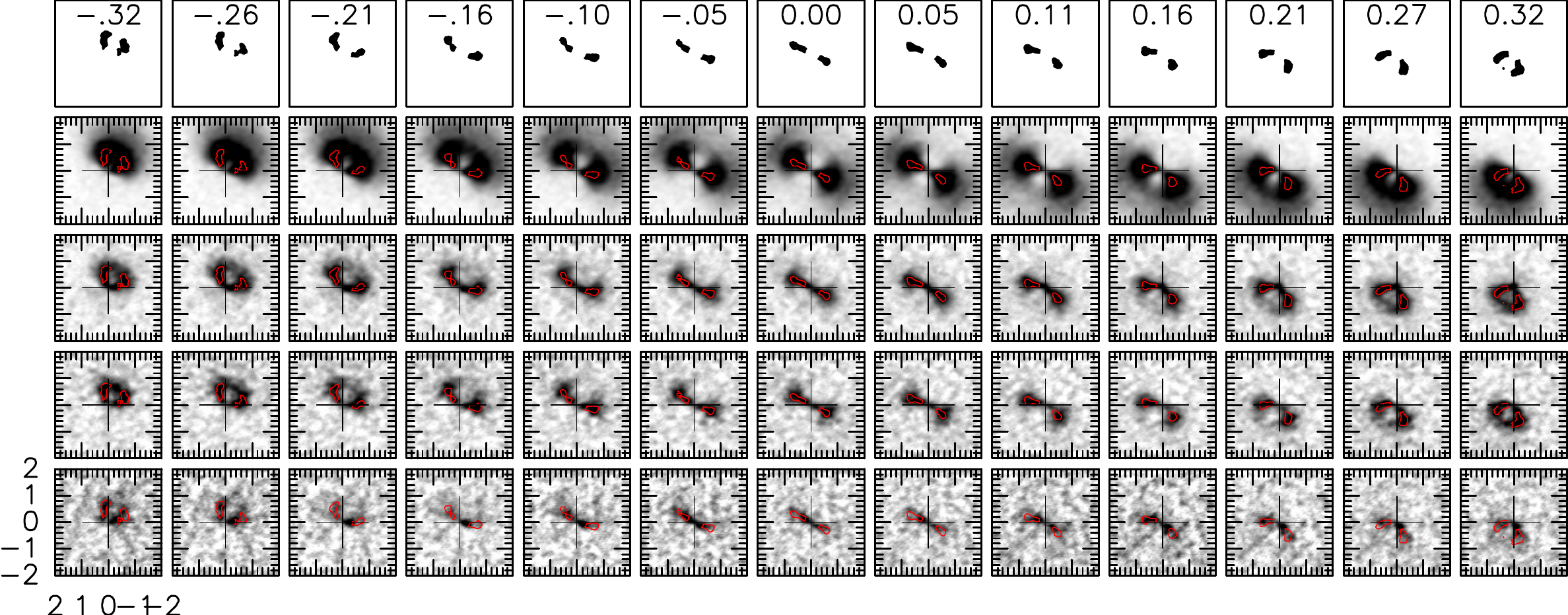}
        \caption{Keplerian masks (top row) overlaid on {(from top to bottom) the HCN (main feature and weak hf line at 345507.455 MHz)}, H\thcn, and HC\fifn, channel maps. The masks (top row) were obtained for a threshold of 40~mJy/beam applied to the weak hf transition at 354507.455~MHz.}
        \label{fig:kmask}
\end{figure*}

\begin{figure*}
        \centering
        \includegraphics[width=\hsize]{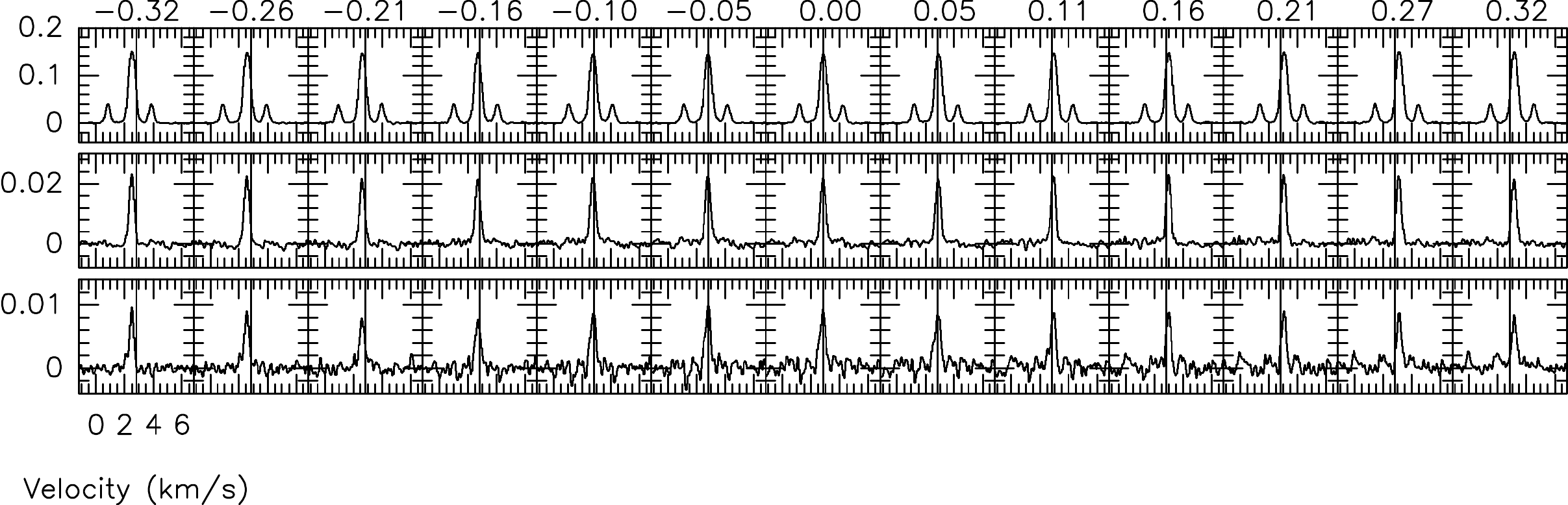}
        \caption{Spatially averaged spectra (Jy/beam) within each Keplerian velocity channel. HCN, H\thcn, and HC\fifn\ are shown from top to bottom. The Keplerian channels are those of \rfig{kmask}.}
        \label{fig:kspectra}
\end{figure*}

In \rfig{kmask} (top row), we show the Keplerian mask defined as explained in the main text. Each mask is overlaid in the channel maps of the three isotopologs, showing the usual dipole-like pattern \citep{beckwith1993}. {Comparing the rows showing the main feature and the weak hf line of HCN, we note} that the extended HCN emission is filtered out in the present study, as is the case of H\thcn, although to a lesser extent. The spectra averaged within each mask, shown in \rfig{kspectra}, were fit with Gaussian profiles, the results of which are summarized in \rtab{kgauss}. The obtained fluxes are then used to derive the \emph{spatially averaged} isotopic ratios listed in \rtab{kratios}.

\subsection{Radial dependence procedure}

\begin{figure*}
        \centering
        \includegraphics[width=0.9\hsize]{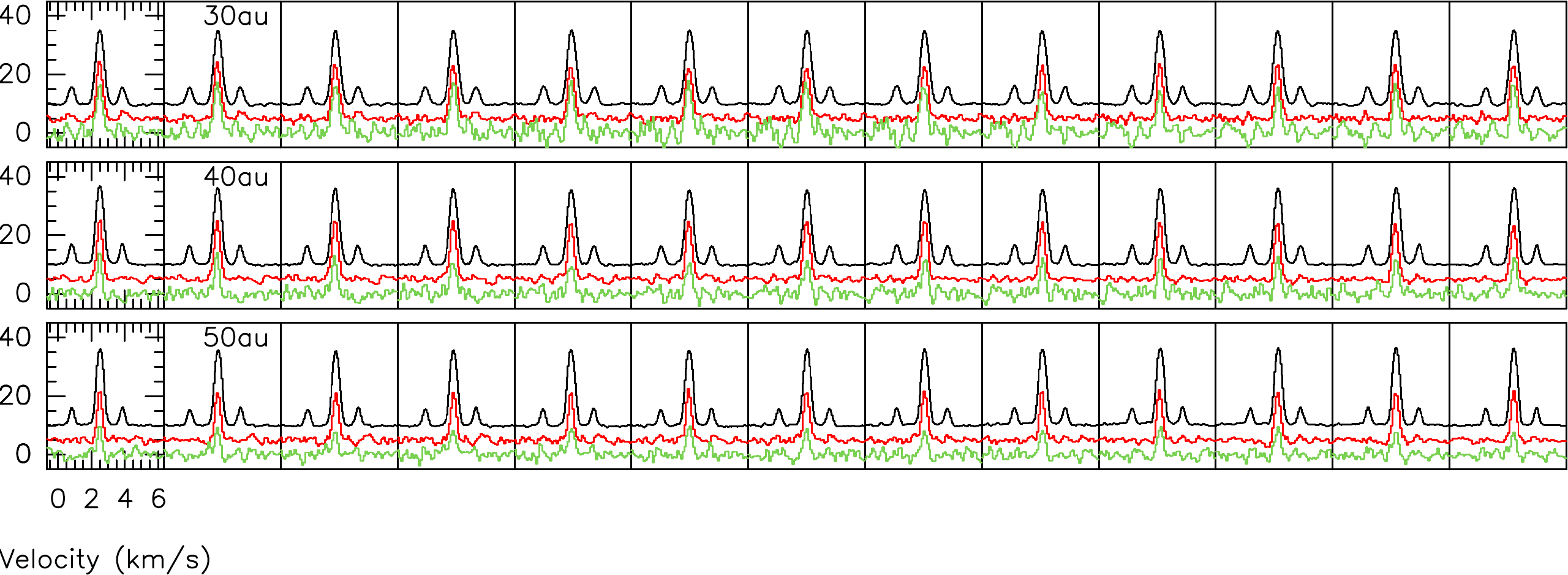}
        \caption{Spectra of HCN (black), H\thcn\ (red, $\times 5$), and HC\fifn\ (green, $\times 10$) averaged within each Keplerian velocity channel (see, e.g., \rtab{kgauss}) and within annuli of constant thickness 5~au, centered at 30, 40, and 50~au (from top to bottom). The specific intensity scale is K using K/(Jy/beam) conversion factors of 178.7, 188.1, and 189.0 for HCN, H\thcn, and HC\fifn, respectively. The spectra are shifted vertically for the sake of legibility.}
        \label{fig:spectra_r_vk}
\end{figure*}

{By combining two masks, one in velocity space (Keplerian mask) and one in the radial direction, we are able to study the radial dependence of the \rr\ isotopic ratio in HCN toward the TW~Hya system. In \rfig{wspectra}, we show the spectra averaged within each annulus, but we concentrate on the central Keplerian channel at 2.83\kms\ (systemic velocity). Figure~\ref{fig:spectra_r_vk} is complementary in that it shows the spectra averaged at three radii, but within all Keplerian masks.}

{As in the radially averaged approach, the spectral line intensity and shape at any radius  do not change appreciably in the various Keplerian velocity channels. In contrast, the dependence on radius is significant, especially for H\thcn\ and most importantly for HC\fifn. As can be seen in \rtab{fwhm} and as mentioned in the main text, all lines become narrower as the radius increases (see \rtab{fwhm}).}


\begin{table*}
        \caption{Results of the Gaussian fits to the spatially averaged spectra of \rfig{kspectra}.}
        \label{tab:kgauss}
        \begin{center}
        \begin{tabular}{r cc cc cc}
                \toprule
                $v_K-v_0^\S$ &\mc{2}{HCN$^a$} & \mc{2}{H\thcn $^b$} & \mc{2}{HC\fifn $^c$} \\
                &
                $W^\S$ & $v^\S$ &
                $W^\S$ & $v^\S$ &
                $W^\S$ & $v^\S$ \\
                \midrule
                -0.32 & 779.2(12.0)& 2.515(1) &   9.8(1) & 2.494(3) & 3.3(1) & 2.491(5) \\
-0.26 & 803.2(13.1)& 2.568(1) &   9.8(2) & 2.541(3) & 3.6(1) & 2.536( 7) \\
-0.21 & 835.1(13.9)& 2.634(1) &   9.6(2) & 2.612(4) & 3.7(2) & 2.611(10) \\
-0.16 & 871.0(12.7)& 2.707(1) & 10.1(2) & 2.698(3) & 3.7(2) & 2.711(11) \\
-0.10 & 884.6(12.2)& 2.753(1) & 10.6(2) & 2.746(4) & 4.1(2) & 2.756(10) \\
-0.05 & 892.4(13.7)& 2.799(1) & 10.9(2) & 2.788(5) & 4.5(2) & 2.784(11) \\
 0.00 & 850.0(14.2)& 2.856(1) & 10.4(2) & 2.834(4) & 4.6(2) & 2.806(11) \\
 0.05 & 838.8(14.1)& 2.899(1) & 10.0(2) & 2.877(4) & 4.5(2) & 2.840(12) \\
 0.11 & 847.7(14.7)& 2.956(1) &   9.7(2) & 2.937(4) & 3.9(2) & 2.921(12) \\
 0.16 & 870.9(16.9)& 3.010(1) &   9.6(2) & 2.988(4) & 3.6(2) & 2.990(10) \\
 0.21 & 881.2(16.1)& 3.067(1) &   9.6(2) & 3.035(3) & 3.5(2) & 3.048( 9) \\
 0.27 & 870.2(15.5)& 3.117(1) &   9.6(2) & 3.080(4) & 3.5(2) & 3.113(10) \\
 0.32 & 859.7(14.2)& 3.166(1) &   9.8(2) & 3.138(3) & 3.7(2) & 3.165(11) \\


                \bottomrule
        \end{tabular}
\end{center}
        \tabnote Statistical uncertainties (at the 1$\sigma$ level) are written in parentheses in units of the last digit.\\
        $^\S$ Keplerian minus systemic velocity (\kms), $W$ is the integrated flux in mJy/beam \kms, and $v$ the central velocity in \kms.\\
        $^a$ For HCN, the integrated intensity is computed as the total flux of the two weak lines divided by the sum of their relative intensities, 4.2\%, while the center velocity is that of the main central component.\\
        $^{b, c}$ The hf lines are not separated, and a single Gaussian was fit to the the main line.
\end{table*}

\begin{table*}
        \caption{Isotopic ratios from the spatially averaged spectra.}
        \label{tab:kratios}
        \centering
        \begin{tabular}{r rl cc cc cc cc cc}
                \toprule
                $v_K-v_0^\S$
                & \mc{2}{$W_{12}^\dagger$}
                & \mc{2}{$W_{13}^\dagger$}
                & \mc{2}{$W_{15}^\dagger$}
                & \mc{2}{HCN/H\thcn$^\ddagger$}
                & \mc{2}{HCN/HC\fifn$^\ddagger$}
                & \mc{2}{H\thcn/HC\fifn$^\ddagger$}\\
                \midrule
                -0.32 & 779.2 & 12.0 &  9.8 & 0.1 & 3.3 & 0.1 & 79.5 & 1.5 & 236.1 &  8.0 & 2.97 & 0.09\\
-0.26 & 803.2 & 13.1 &  9.8 & 0.2 & 3.6 & 0.1 & 82.0 & 2.1 & 223.1 &  7.2 & 2.72 & 0.09\\
-0.21 & 835.1 & 13.9 &  9.6 & 0.2 & 3.7 & 0.2 & 87.0 & 2.3 & 225.7 & 12.8& 2.59 & 0.15\\
-0.16 & 871.0 & 12.7 & 10.1& 0.2 & 3.7 & 0.2 & 86.2 & 2.1 & 235.4 & 13.2& 2.73 & 0.16\\
-0.10 & 884.6 & 12.2 & 10.6& 0.2 & 4.1 & 0.2 & 83.5 & 2.0 & 215.8 & 10.9& 2.59 & 0.14\\
-0.05 & 892.4 & 13.7 & 10.9& 0.2 & 4.5 & 0.2 & 81.9 & 2.0 & 198.3 &  9.3 & 2.42 & 0.12\\
 0.00 & 850.0 & 14.2 & 10.4& 0.2 & 4.6 & 0.2 & 81.7 & 2.1 & 184.8 &  8.6 & 2.26 & 0.11\\
 0.05 & 838.8 & 14.1 & 10.0& 0.2 & 4.5 & 0.2 & 83.9 & 2.2 & 186.4 &  8.9 & 2.22 & 0.11\\
 0.10 & 847.7 & 14.7 &  9.7 & 0.2 & 3.9 & 0.2 & 87.4 & 2.4 & 217.4 & 11.8& 2.49 & 0.14\\
 0.16 & 870.9 & 16.9 &  9.6 & 0.2 & 3.6 & 0.2 & 90.7 & 2.6 & 241.9 & 14.2& 2.67 & 0.16\\
 0.21 & 881.2 & 16.1 &  9.6 & 0.2 & 3.5 & 0.2 & 91.8 & 2.5 & 251.8 & 15.1& 2.74 & 0.17\\
 0.27 & 870.2 & 15.5 &  9.6 & 0.2 & 3.5 & 0.2 & 90.6 & 2.5 & 248.6 & 14.9& 2.74 & 0.17\\
 0.32 & 859.7 & 14.2 &  9.8 & 0.2 & 3.7 & 0.2 & 87.7 & 2.3 & 232.4 & 13.1& 2.65 & 0.15\\


                \bottomrule
        \end{tabular}
        \tabnote $\S$: Keplerian minus systemic velocity (\kms) is that of the hf line of HCN at 354507.455~MHz. $\dag$: $W_{12}$, $W_{13}$, and $W_{15}$ are the integrated fluxes (mJy/beam \kms) derived from Gaussian fitting (see also \rtab{kgauss}). $\ddagger$: integrated line flux ratios $W_{12}/W_{13}$, $W_{12}/W_{15}$, and $W_{13}/W_{15}$. All uncertainties are statistical and at the 1$\sigma$ level.
\end{table*}

\subsection{Radially dependent line width}
Gaussian fits to Keplerian-filtered spectra averaged within concentric annuli of thickness 0.1\arcsec\ were performed in order to measure the dependence of the FWHM of the HCN, H\thcn, and HC\fifn\ lines. The results are summarized in \rtab{fwhm}. They are used to infer an upper limit on the kinetic temperature following \hbmk\ and \cite{teague2016}.

\begin{table}
        \caption{Radial dependence of the line width (FWHM, \kms) of HCN, H\thcn, and HC\fifn.}
        \label{tab:fwhm}
        \centering
        \begin{tabular}{c c@{$\pm$}c c@{$\pm$}c c@{$\pm$}c}
                \toprule
                $r$ (\arcsec)
                & \mc{2}{HCN}
                & \mc{2}{H\thcn}
                & \mc{2}{HC\fifn} \\
                \midrule
                0.3 & 0.954 & 0.183 & 0.832 & 0.101 & 1.231 & 0.140\\
                0.4 & 0.614 & 0.060 & 0.603 & 0.038 & 0.687 & 0.043\\
                0.5 & 0.461 & 0.020 & 0.458 & 0.034 & 0.479 & 0.071\\
                0.6 & 0.389 & 0.018 & 0.382 & 0.020 & 0.355 & 0.036\\
                0.7 & 0.370 & 0.013 & 0.369 & 0.020 & 0.316 & 0.048\\
                0.8 & 0.354 & 0.014 & 0.371 & 0.016 & 0.334 & 0.053\\
                0.9 & 0.345 & 0.016 & 0.367 & 0.030 & 0.354 & 0.058\\
                \bottomrule
        \end{tabular}
        \tabnote Spectra were averaged in annuli of thickness 0.1\arcsec. $r$ is the outer radius of each annulus.
\end{table}

\section{Single excitation temperature assumption}
\label{app:tex}

The main assumption in our analysis is that of a single excitation temperature characterizing the weak hf lines of HCN and the overlapping hf lines of H\thcn\ and of HC\fifn. A thorough investigation of this assumption can only be made with two-dimensional radiative transfer taking into account the overlap of hf transitions in the level population evaluation, which in turn requires a complete physical and kinematic structure of the disk. This type of analysis has been made in the context of pre-stellar starless cores by \cite{magalhaes2018a}, in spherical geometry, with the 1D ALICO model \citep{daniel2008}.

\begin{figure}
        \includegraphics[width=\hsize]{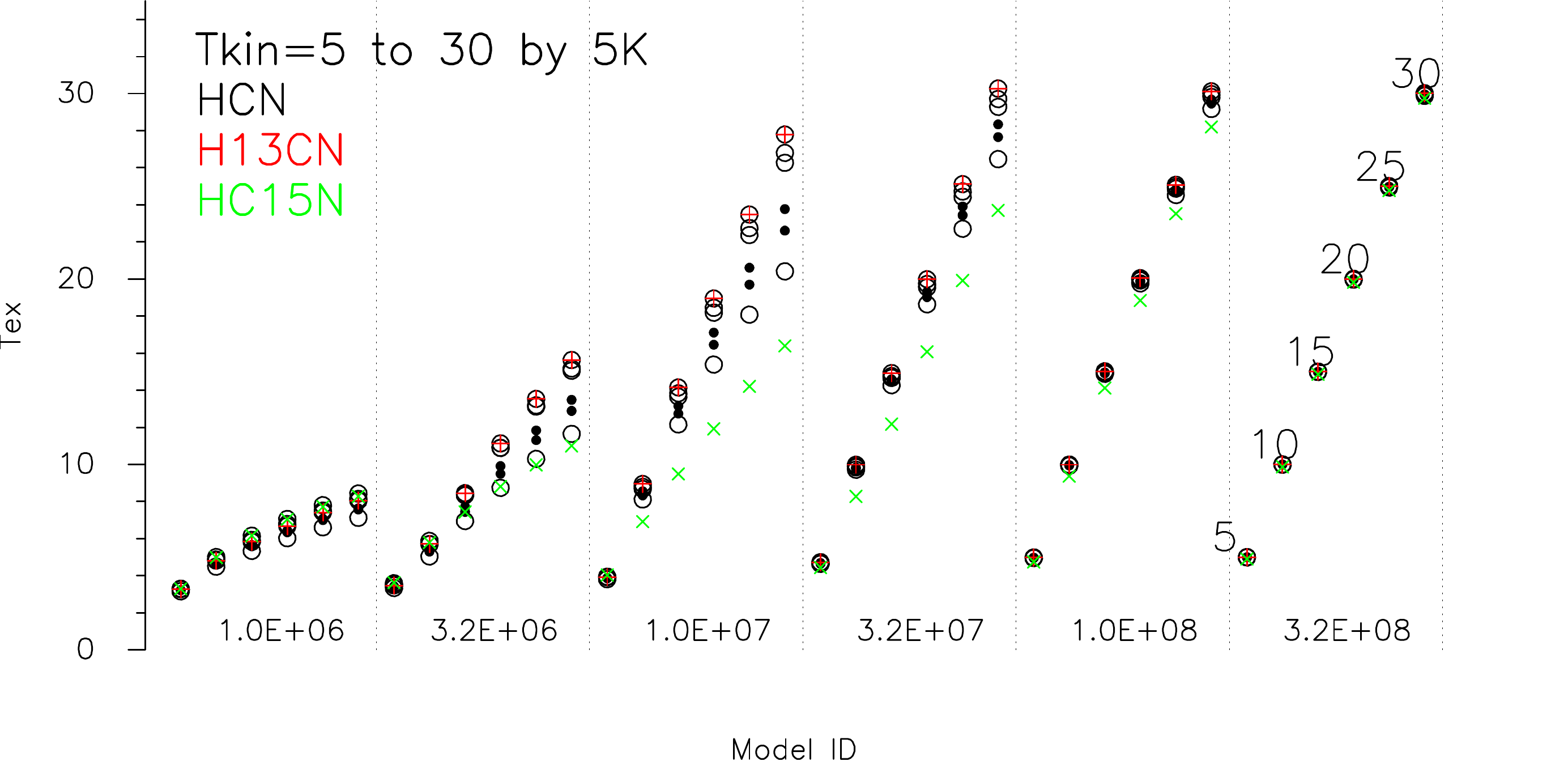}
        \caption{Excitation temperature of the six hf lines of HCN (see \rtab{spectro}) and the strongest hf lines of H\thcn\ (red) and HC\fifn\ (green). The four overlapping hf lines of HCN are shown as open circles and the two weakest resolved lines with filled dots.}
        \label{fig:tex}
\end{figure}

In the present context, we were unable to make detailed calculations like this. Instead, we used our 1D ALICO code to investigate the density and kinetic gas temperature at which departures from the single excitation temperature assumption may be obtained. To do so, the level populations were computed while treating the effect of hf line overlap in the excitation of HCN, H\thcn, and HC\fifn. Our grid of models encompasses a range of \hh\ density (\dix{6} to \dix{11}\ccc) and kinetic temperature (from 5 to 30 K) relevant to the TW~Hya orbiting disk. The results of these calculations in terms of \texc\ are shown in \rfig{tex}. There are significant departures at densities between \dix{6}\ccc\ and \dix{8}\ccc. At higher density, all lines are thermalized (models at \nhh\ higher than \dix{9}\ccc\ are not shown). From the generic chemical models of \cite{visser2018}, it appears that HCN is concentrated at radii $r=5-50$~au and scale heights $z/r=0.2$ to 0.3, where the gas kinetic temperature is $\approx 200$K and the density in the range \dix{6} to \dix{8}\ccc, which would indicate that the single excitation temperature assumption fails.

On the other hand, the FWHM of the HCN, H\thcn, and HC\fifn\ lines at a radius of 0.6\arcsec\ ($\approx$35~au) and beyond is $\approx 0.35$\kms\ (see \rtab{fwhm}), placing a conservative upper limit on the kinetic temperature of \citep{teague2016}
\begin{equation}
(FWHM/2.35)^2\times\mu \mh/(2k_B)\approx35 \rm\,K.
\end{equation}
Referring to the tailored TW~Hya model of \cite{vanthoff2017}, this upper limit, applied to the relevant range of radii, indicates that the density is higher than \dix{9}\ccc, in a regime where lines should be thermalized. This validates our single excitation temperature assumption. Our upper limit on the kinetic temperature is also well below the gas temperature of $\sim200$~K where the HCN abundance is the highest in the models of \cite{visser2018}, although we note that these models are generic and are not tailored to TW~Hya.

\end{document}